\documentclass{revtex4}
\usepackage{graphicx}
\usepackage{color}
\usepackage{graphics}
\begin{document}
\title{La Orden del Caballero Bondi\\ \textcolor{blue}{The Order of Sir Bondi}}
\author{W. Barreto}
\email{wbarreto@ula.ve}
\affiliation{Centro de F\'\i sica Fundamental, Facultad de Ciencias,
\\Universidad de Los Andes, M\'erida, Venezuela}
\maketitle
\begin{center}Resumen\end{center}
Donde narro la relaci\'on directa e indirecta del
f\'\i sico brit\'anico de origen austr\'\i aco, Sir Hermann Bondi,
con la f\'\i sica venezolana. Hago especial \'enfasis
en sus cualidades humanas m\'as notables, nobleza y humildad,
aparte de su importante contribuci\'on a la Relatividad General y 
a la organizaci\'on de la ciencia europea. Auto-declarado como un leal seguidor
de la filosof\'\i a  popperiana, Bondi defiende a la ``dura y sucia'' f\'\i sica
en contraposici\'on a la ``inma\-culada y bella'' f\'\i sica.
Centro atenci\'on en dos piedras angulares de la obra de Bondi,
seguidas a profundidad por Luis Herrera y disc\'\i pulos en Venezuela.

\textcolor{blue}{\begin{center}{Abstract}\end{center}
The direct and indirect relationship
between the British physicist originally from Austria, Sir
Hermann Bondi,  and Venezuelan physics are presented. Special emphasis is made
of his more remarkable human qualities, nobility and humbleness, besides
his important contribution to General Relativity and to the
European science agencies. Self--declared as a staunch disciple of
Popperian philosophy, Bondi defended the ``hard and dirty'' physics against
the ``inmaculate and beauty'' physics. Attention is focused on two cornerstones of Bondi's work, 
followed thoroughly by Luis Herrera and disciples in Venezuela.}

---------------------------------------------------------------------------------------------------------------------------------------------------
\vskip 0.25cm

{La caballer\'\i a est\'a asociada a la nobleza}; cierta actitud incorrup\-tible 
siempre a prueba en el crisol del tiempo, imbatible bajo cualquier circunstancia.
As\'{\i} era Sir Hermann Bondi, noble.
As\'{\i} son Luis Herrera y
Jeffrey Winicour. Como suelen decir las abuelas ``ese material, ya no sale m\'as''.
Al parecer, Bondi tambi\'en
era humilde. Muy lejos de padecer el s\'\i ndrome de las {\it vedettes} en el firmamento de la ciencia global.
Algo bastante poco usual, considerando que Bondi es referenciado por la Enciclopedia Brit\'anica y que fue Director de la Agencia para la Investigaci\'on Aeroespacial Europea, organizaci\'on equivalente a la NASA.  Hubiera sido estupendo conocerle en persona... y pensar que estuve cerca. A continuaci\'on elaborar\'e una breve semblanza de Hermann Bondi, basada en pocas
an\'ecdotas y escritos suyos, sustantivos e iluminadores. Luego abordar\'e una revisi\'on no rigurosa de dos trabajos de la obra de Bondi. Estos trabajos han sido objeto de estudio en Venezuela por Luis Herrera y sus disc\'\i pulos.

\vskip 0.25cm
\textcolor{blue}{Knighthood and nobility are associated; certain incorruptible attitude always under test on the cradle of time, unbeatable under any circumstance. That was Sir Hermann Bondi, noble, that are  Luis Herrera and Jeffrey Winicour. How use to say grandmothers ``that material, is not coming anymore''. Seems to be that Bondi also was humble, very far from the syndrome of vedettes in the heavens of global science. Something too little usual, considering that Bondi is referenced by the Britannic Encyclopedia and he was Director of the European Space Research Organization, equivalent to NASA. I would have liked met him... I was this close. In what follows I elaborate a biographical sketch of Hermann Bondi, based on few anecdotes and written of
him, substantives and illuminating. Then, a non exhaustive revision of two seminal papers of Bondi's work is made. These articles have been extensively studied by Luis Herrera and disciples in Venezuela.}

-----------------------
\vskip 0.25cm
Desde los a\~nos cincuenta se realizan cursos de verano sobre Relatividad en distintas partes del
mundo. Uno de esos cursos se realiz\'o en el King's College, en Londres.
A \'este asisti\'o el joven americano Jeffrey Winicour, doctorado
por Syracuse bajo la direcci\'on de Peter Bergmann, quien trabaj\'o con Einstein. Winicour es fundador del Grupo de Relativistas
de Pittsburgh, junto con Ezra Ted Newman y Allen Janis. Trabajar con Winicour ha sido una fortuna cuando adem\'as se tiene en com\'un 
el gusto por montar bicicleta y  fumar puros. En una de esas charlas
de sobre mesa en el porche de su casa y bajo los efluvios del tabaco, Jeff coment\'o
que hab\'\i a conocido a Bondi (en cuya obra Winicour bas\'o la suya). Record\'o algo as\'{\i}:{\it ``Esper\'abamos por la clase siguiente
y un se\~nor se apareci\'o a la hora, justo antes de la clase, a borrar el pizarr\'on. Todos est\'abamos
algo distra\'\i dos esperando al profesor. El se\~nor de la limpieza
comenz\'o a dirigirse al auditorio, supimos de inmediato que era Bondi. Est\'abamos desconcertados, llevaba la camisa arre\-mangada y} --como una guayabera-- {\it por fuera}. {\it Quedamos atrapados en el acto por su persona\-lidad afable y por su profundidad con el fluir del discurso de la clase magistral.''} Qued\'o callado entre volutas. Winicour dijo entonces que est\'abamos en una especie de ``recesi\'on relativista''. Hay pocos relativistas num\'ericos y una vez formados se han visto atra\'\i dos por el sector financiero o la industria.
Hacen falta caballeros para la causa, para entender cada vez m\'as. 
Seguro que la comunidad ha crecido, sin duda, pero no
necesariamente la calidad en la misma proporci\'on. Tambi\'en es claro que se ha profundizado, pero la
industria de art\'\i culos sobre los aspectos m\'as ``rentables'' de la Rela\-tividad General, como m\'as o menos
dice Luis He\-rrera, francamente do\-mina el medio.

\vskip 0.25cm
\textcolor{blue}{From fifties summer schools on Relativity are organized in the entire world. One of them was in London King's College. It was attended by the young american Jeffrey Winicour, who got his PhD in Syracuse University under supervision of Peter Bergmann, who works with Einstein. Winicour is founder of the Pittsburgh Relativity Group with Ezra Ted Newman and Allen Janis. Working with Winicour has been fortunate, if one have additionally in common the pleasure of ride bikes and smoke cigars. In one of these talk after dinner, sitting on his porch's house, immerse in tobacco effluvia, Jeff told me he met Bondi (in whose work Winicour based his own). He remembered something like this: ``{\it We were waiting for the next class and one man appeared on time to erase
the blackboard. Everyone was wandering around waiting for the professor. The cleaning man begun to talk to the audience... we knew immediately he was Bondi. We were disconcerted, He wore the shirt with rolled up sleeves and} --like a shirt for guavas-- {\it loosed. We were trapped {\it ipso facto} because his affable personality and his deep and fluent speech for the magistral class.}''
He kept wordless among curls and vortices. He said then we are in a kind of ``relativistic recession''. There are few numeric relativistic practitioners and once formed they have been
attracted by the financial sector or industry. Knights are needed, to understand more and more. The community has grown, without doubt, but not necessarily in quality in the same proportion. Clearly we have a better view, but the industry of papers on rentable aspects of General Relativity, as more or less Luis Herrera said, frankly fill the environment.}

-----------------------
\vskip 0.25cm
En el obituario de la BBC de Londres sobre Bondi, se lee que despu\'es de su desacreditado mode\-lo cosmol\'ogico del estado estacionario, Bondi se dedic\'o a estudiar la f\'\i sica de los agujeros negros. Esto no es totalmente cierto. Aunque public\'o
algunos trabajos sobre agujeros negros, la mayor contribuci\'on de Bondi a la Rela\-tividad Ge\-neral fue
 el estudio y comprensi\'on de la radiaci\'on gravitacional.
 El modelo del estado estacionario propuesto en coautor\'\i a con Hoyle y Gold,
visto en retrospectiva y en su contexto hist\'orico, resulta atractivo y audaz \cite{bhg}. Por cierto, Bondi conoce a
Thomas Gold en una reclusi\'on preventiva en Canad\'a. Como inmigrantes
austr\'\i acos en el Reino Unido se les investig\'o al comenzar la segunda guerra mundial;
luego ambos se reunieron con Fred Hoyle en un proyecto
sobre radares en el Almirantazgo Brit\'anico. Su visi\'on sobre la acreci\'on
condujo a Hawking a formular la radiaci\'on proveniente de agujeros negros. Bondi fue nombrado Caballero del Reino
Unido en 1973.

\vskip 0.25cm
\textcolor{blue}{In the London's BBC obituary on Bondi it reads that after his discredited steady state cosmological model, he dedicated to the study of black holes physics. This is not true
at all. Although he published some works on black holes, Bondi's main contribution to General Relativity was the study and comprehension of the gravitational radiation. The proposed steady state model coauthored with Hoyle and Gold, seen retrospectively and in its historical context, is attractive and audacious \cite{bhg}. By the way, Bondi met Thomas Gold in a preventive reclusion in Canada. As Austrian immigrants in United Kingdom they were investigated beginning the second war world; later both met Fred Hoyle in a project about radars in the British Almiranty.
His vision on accretion leads Hawking to propose that black holes radiates. Bondi was named Knight of the United Kingdom in 1973.}

-----------------------
\vskip 0.25cm
La primera vez que supe de Bondi fue por el a\~no 1987, cuando mi tutor y sensei Luis Herrera se\~nal\'o, para una revisi\'on profunda, un art\'\i culo \cite{hjr} que representaba en cierto modo la continuaci\'on
de otro art\'\i culo de 1964 cuyo autor es Hermann Bondi. M\'as adelante regresar\'e con detalle
sobre este \'ultimo trabajo cuya riqueza no se ha explotado tanto como
el trabajo sobre radiaci\'on gravitacional del a\~no 1962. Cabe mencionar que
Luis Herrera es graduado summa cum laude por la Universidad Patricio Lumumba de Mosc\'u  y luego docto\-rado en el Instituto Henri Poincar\'e de Par\'\i s, bajo la direcci\'on de Achille Papapetrou, quien a su vez y a su tiempo trabaj\'o con Erwin Schr\"odinger.
\vskip 0.25cm
\textcolor{blue}{The first reference that I had to Bondi was about 1987, when my advisor and Sensei Luis Herrera pointed out, for a deep revision, one paper \cite{hjr}  which in some way represented the continuation of another paper published in 1964 whose author is Hermann Bondi. Below I will come back with some detail to this Bondi's work, whose richness has not been exploited enough as his celebrated 1962 paper on gravitational radiation. Parenthetically, Luis Herrera is graduated summa cum laude by Moscu's Patricio Lumumba University and got his PhD by Paris's Henry Poincar\'e Institute, under the direction of Achille Papapetrou, whom at his time and epoch worked with Erwin Schr\"odinger.}

-----------------------
\vskip 0.25cm
En 1992  Bondi public\'o un trabajo sobre colapso gravi\-tacional en distribuciones anis\'otropas \cite{b92}. Luis Herrera,
aunque ya hab\'\i a publicado un notable n\'umero de trabajos sobre el tema,
no figuraba referenciado en el art\'\i culo de Bondi. Tom\'o un sobre manila y adjunt\'o sin explicaci\'on
al destinatario un cartapacio de art\'\i culos sobre anisotrop\'\i a en distribuciones esf\'ericas y su efecto sobre el colapso gravitacional \cite{h_varios}. A vuelta de correo recibi\'o una misiva fechada el 26 de enero de 1993, que traduzco \'\i ntegra y libremente a continuaci\'on: 
\vskip 0.25cm
{\it Estimado Dr Herrera,

 Por favor, acepte mis profundas disculpas por la fa\-llida referencia en mi art\'\i culo (M.N.R.A.S. 259. 365-368, 1992)
al extenso trabajo que usted y sus colegas han rea\-lizado sobre esferas anis\'otropas en Relatividad Ge\-neral. Todav\'\i a 
no logro entender c\'omo pudo suceder esto, por cuanto pens\'e hab\'\i a realizado una b\'usqueda adecuada en la que
hab\'\i a involucrado a un muy reconocido colega. Sin embargo, la responsabilidad por esta desafortunada omisi\'on
es completamente m\'\i a.

De la lectura de los art\'\i culos que usted tan amablemente me ha enviado me parece que la
optimizaci\'on particular que emprend\'{\i} no repite cualquiera de sus trabajos, aunque en el
de 1980 con Cosenza, Esculpi y Witten su acotaci\'on que sigue a la ecuaci\'on (21) se\~nala hacia
el \'area que explor\'e. Mi mortificaci\'on fue mayor por sus generosas referencias a mi trabajo previo.

Estoy escribiendo al Editor de M.N. pregunt\'andole si me permite publicar una nota de disculpas, quiz\'as como
un erratum. Para asegurarme que lo he hecho bien, esta vez me gustar\'\i a enviarle un fax con la nota, de tal
forma que usted pueda se\~nalarme cualquier error ?`podr\'\i a usted en consecuencia enviarme su n\'umero de fax?

Por favor, transmita mis disculpas tambi\'en a sus colegas. Conf\'\i o en que mi esperada nota en M.N. pueda ser
vista por ellos a su debido tiempo.

Una vez mas con mis disculpas,  

Atentamente,

Hermann Bondi}
\vskip 0.25cm

Desde entonces Bondi no faltaba en casa de los Herrera--Di Prisco con una tarjeta cada Navidad.
\vskip 0.25cm
\textcolor{blue}{In 1992 Bondi published one work on gravitational collapse of anisotropic
distributions \cite{b92}. Luis Herrera, having published a remarkable number of papers
on the subject, did not appear referenced in Bondi's paper. Took a Manila's paper envelope and attached without any explanation to the receiver a bundle of papers on anisotropy in spherical distributions and its effect on gravitational collapse \cite{h_varios}. To mailing reply he received the following letter dated January 26, 1993:}

\textcolor{blue}{{\it Dear Dr Herrera,}}

\textcolor{blue}{\it Please accept my profound apologies for failing to refer in my paper (M.N.R.A.S. 259, 365-368, 1992) to the extensive work you and your colleagues have done on anisotropic spheres in General Relativity. I am still at a loss to understand how this came about, for I thought I had searched adequately and had also involved a usually very knowledgeable colleague. However, the responsibility for this unfortunate omission is entirely mine.}

\textcolor{blue}{\it From my reading of the papers you so kindly sent me it appears to me that the particular optimisation I undertook did not repeat any of your workings, though in your 1980 paper with Cosenza, Esculpi and Witten your remark following equation (21) points towards the area I looked at. My mortification is made the greater by your many generous references to my earlier work.}

\textcolor{blue}{\it I am writing to the Editor of M.N. asking him  to allow me to publish a note of apology, perhaps as an erratum. To make sure I get it right this time I would like to fax it to you so that you can point out any errors to me. Could you therefore please let me have your fax number?}

\textcolor{blue}{\it Please convey my apologies also to your colleagues. I trust my hoped-for note in M.N. should reach them all due course.}

\textcolor{blue}{\it Again apologetically}

\textcolor{blue}{\it Yours sincerely}

\textcolor{blue}{\it Hermann Bondi}
\vskip 0.25cm
\textcolor{blue}{From then on Bondi sent a card to Herrera--Di Prisco's family each Christmas holidays.}

-----------------------
\vskip 0.25cm
En mayo del a\~no 2000 Bondi visita a Herrera en Salamanca, de lo cual es testimonio fotogr\'afico la figura 1. Para entonces su Parkinson estaba muy avanzado y
tomaba unas grageas fulminantes que de vez en cuando lo desconectaban de la l\'\i nea central del mundo. Pero en su charla
magistral fue l\'ucido, afable, humilde, contundente y locuaz, como siempre. Un caballero de la luz que ilumin\'o
la gravitaci\'on. Bondi fue condecorado por las autoridades de la Universidad de Salamanca.
\begin{figure}[!ht]
\includegraphics[width=7in]{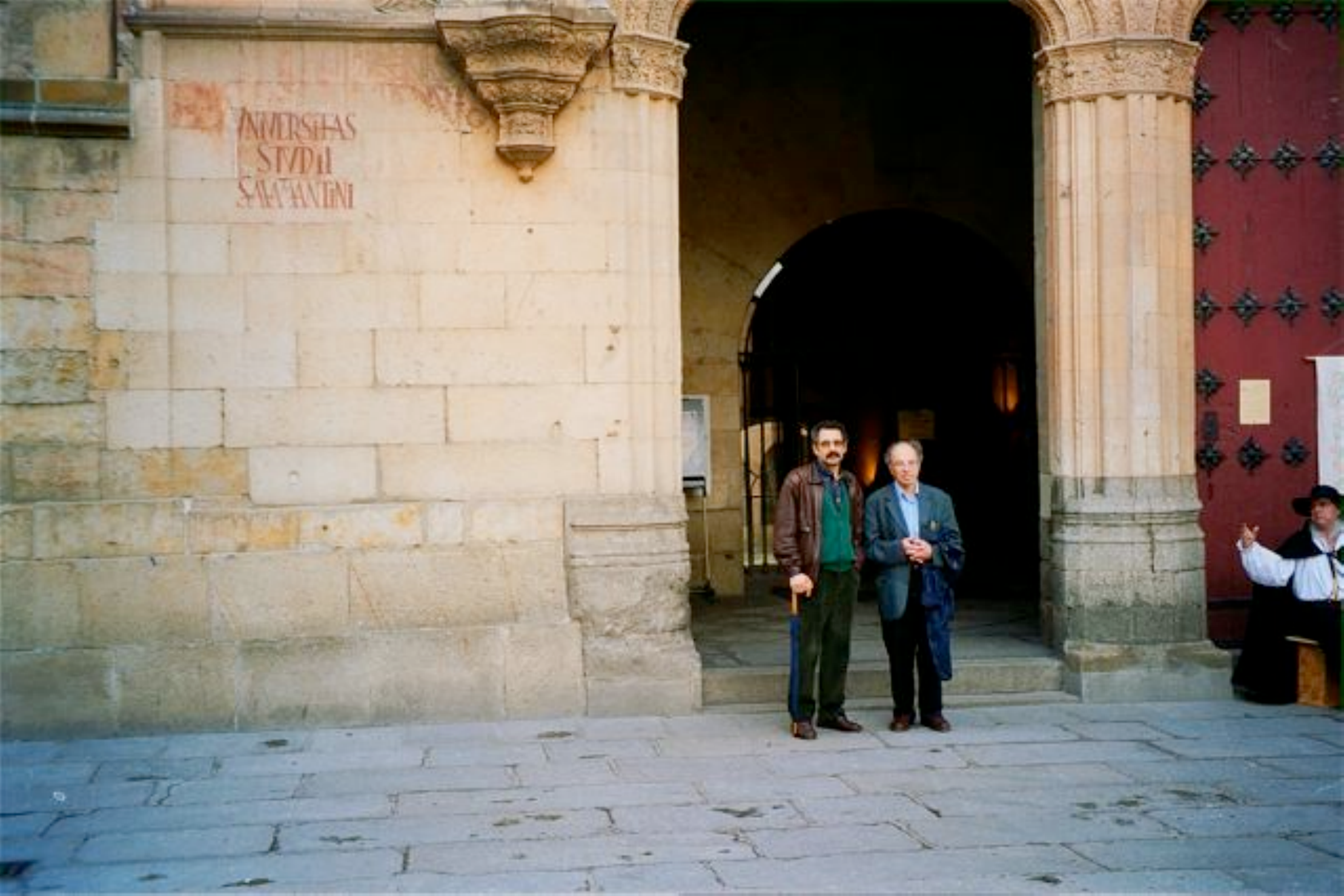}
\caption{Herrera y Bondi en Salamanca, mayo del 2000.}
\label{fig:LH_HB}
\end{figure}
De vuelta en Londres, Bondi hace referencia a su visita y sobre todo al trabajo
de Luis Herrera. 
En una carta fechada el 17 de mayo del a\~no 2000, muestra su genuino inter\'es por la f\'\i sica:
\vskip 0.25cm
{\it Estimado Luis,

La vida aqu\'{\i} desde mi regreso ha sido absolutamente fren\'etica; por \'eso el retraso
en escribirte esta carta. Los d\'\i as que pas\'e en Salamanca fueron los m\'as agradables
y gratos para m\'{\i} y deseo agradecerte a t\'{\i}, a tu esposa, a Jes\'us Mart\'\i n y a los
dem\'as (incluyendo a tus hijos) por tan c\'alidos, placenteros e interesantes momentos.

El retorno al aeropuerto de Madrid fue r\'apido, con una breve `parada de confort' en un peque\~no
caf\'e del camino. Mi vuelo estuvo  bien, con un ligero retraso. Pas\'e el tiempo en el
aeropuerto comiendo y leyendo, especialmente disfrutando tus art\'\i culos. El uso de un 
gir\'oscopo} --Bondi se refiere a la deteccci\'on de radiaci\'on gravitacional-- {\it representa una forma eficiente para aclarar situaciones relativistas en
un escenario relevante. Qued\'e particularmente complacido con tu art\'\i culo sobre `relajaci\'on',
como he insistido repetidamente en cuanto a la necesidad de considerar relaciones constitutivas realistas junto con las ecuaciones
de campo.  Esto debe requerir una dependencia temporal para la respuesta del material a los 
esfuerzos gravitacionales. S\'olo de esta forma la fricci\'on de marea puede entrar en el an\'alisis.
Para el pulsar binario de Taylor es importante mostrar que la fricci\'on de marea tiene mucho menos
influencia sobre la evoluci\'on de la \'orbita que la reacci\'on radiativa. No estoy al tanto si ya
esto est\'a establecido satisfactoriamente. [Pudieras estar interesado en saber que hace cincuenta a\~nos publiqu\'e
un art\'\i culo sobre el efecto l\'\i mite de la velocicad del sonido en el oc\'eano sobre la rapidez de las
olas.]}

{\it Ayer recib\'{\i} la feliz noticia de que mi art\'\i culo sobre ondas gravitacionales (del
cual hiciste una fotocopia) ha sido aceptado por la Sociedad Real, con un arbitro estusias\-ta con el
trabajo y otros que lo aprobaron en tonos m\'as moderados.}

{\it Otra vez muchas gracias: Me dieron momentos muy placenteros y s\'olo puedo esperar que ustedes
los hayan disfrutado tambi\'en.

Con mis mejores deseos para todos,

Hermann}
\vskip 0.25 cm

De esta \'ultima carta sorprende la actividad y lucidez de Bondi a sus 81 a\~nos.
Pero sobre todo es grati\-ficante el
entusiasmo y la alegr\'\i a que manifest\'o por un art\'\i culo m\'as en el zurr\'on.
Esto permite se\~nalar lo siguiente: el \'ultimo trabajo publicado es el
m\'as importante. Pero esto no es cierto en la obra singular de Bondi. Como se ver\'a, entre los a\~nos 1962
y 1964 public\'o dos joyas de la literatura en Relatividad General.
\vskip 0.25cm
\textcolor{blue}{By May, 2000, Bondi visited Luis Herrera at Salamanca, of which is photographic testimony the figure 1. By then his Parkinson was very advanced and he had some fulminant pills, from time to time, which disconnected him from the central world line. But, his talk was lucid, affable, humble, straighten and of easy speech, as ever. One light's Knight who illuminated gravitation. Bondi was honored by the Salamanca's University rulers. Once returned to London, Bondi makes mention to his visit and specially to Luis Herrera's work. In a letter dated May 17, 2000, he showed his genuine interest in physics:}

\textcolor{blue}{\it Dear Luis,}

\textcolor{blue}{\it Life here has been absolutely hectic since my return; hence the delay in writing this letter. The days that I spent in Salamanca were most agreeable and enjoyable for  me and I want to thank you, your wife, Jesus Martin and the others (including of course your boys) most warmly for this most interesting and pleasant time.}

\textcolor{blue}{\it The return drive to Madrid airport was quick, with a brief `confort stop' at a little caf\'e on the road. My flight was fine, with only a slight delay. I spent my time at the airport eating and reading, particularly enjoying your papers. Using a gyroscope is a very neat way of clarifying relativistic situations in a physically significant way. I was particularly thrilled by your `relaxation' paper, as I have repeatedly stressed the need to get realistic constitutive relations into the field equations. This must require a time dependence of the response of the material system to gravitational stresses. Only in such a manner can tidal friction be brought into the analysis. For Taylor's binary pulsar it is important to show that tidal friction has much less influence on the evolution of the orbit than the radiative reaction. I do not know whether this has been settled satisfactorily. [You may be interested that half a century ago or so I published a paper on the limiting effect of the finite velocity of sound in the ocean on the speed of surface waves.]}

\textcolor{blue}{\it Yesterday I got the happy news that my gravitational wave paper (of which you took a photo copy) has been accepted by the Royal Society, with one referee enthusiastic about it and the other approving in more measured tones.}

\textcolor{blue}{\it Again many thanks: You gave me a lovely time and I can only hope you all enjoyed it too.}

\textcolor{blue}{\it With kind regards to all}

\textcolor{blue}{\it Yours sincerely}

\textcolor{blue}{\it Hermann}

\vskip 0.25cm
\textcolor{blue}{From this last letter surprised me the activity and lucidness of Bondi in his 81. But it is good to know the enthusiasm and happiness showed because one more paper in the leather bag. This let us point out: the last published paper is the most important. But, this is not true in the singular Bondi's work. As it will be seen, between 1962 and 1964 he published two
jewelry literature in General Relativity.}

-----------------------
\vskip 0.25cm
Bondi y Herrera mantuvieron comunicaci\'on durante unos diez a\~nos. Luego de la visita a Herrera en Salamanca ocurri\'o un incidente que vale la pena mencionar y que, una vez mas, nos ayuda a entender el talante noble de Sir Hermann Bondi. Luis Herrera por ser extranjero fue impedido de ocupar una plaza permanente en la Universidad de Salamanca. Bondi indignado intent\'o publi\-car una carta de protesta en {\it Nature}. El Editor evadi\'o a Bondi,  indign\'andose \'este a\'un m\'as. Pidi\'o disculpas a Herrera insistentemente y mostr\'o preocupaci\'on por el regreso de la familia a Caracas.
\vskip 0.25cm
\textcolor{blue}{Bondi and Herrera kept in touch about ten years. After his visit to Herrera at
Salamanca happened an incident which it is worth the pain to mention and, once more, help us to understand the Sir Hermann Bondi nobility and willingness. Luis Herrera as a foreign was excluded to get a tenure at Salamanca University. Outraged Bondi try to publish a protesting letter in {\it Nature}. The Editor evaded Bondi, and he became angry. He begged apologies repetitively to Herrera and showed preoccupied because the family's return to Caracas.}

-----------------------
\vskip 0.25cm
Es ejemplarizante que Bondi, habiendo ocupado cargos de alt\'\i simo rango en Europa y en Inglaterra, mantuviera 
un vivo inter\'es por la investigaci\'on durante toda su ca\-rrera y que a la vez fuera notable y abrumadoramente humilde.
Es indudable la enorme estatura de Bondi, quien entreg\'o su vida al servicio de la ciencia. Fue influyente en la din\'amica del mundo real, matem\'atico, astrof\'\i sico, cosm\'ologo, pero sobre todo fue un Caballero de la Ciencia. Publi\-c\'o trabajos
con figuras legendarias como F. Pirani \cite{bp}, D. Sciama \cite{bgs} y W. Rindler \cite{br}.
\vskip 0.25cm
\textcolor{blue}{It is instructive that Bondi, having occupied positions of high rank in Europe and England, kept a vivid interest for research in the course of his career and at the same time he was remarkable and overwhelmingly humble. There is no doubt about his enormous stature, who dedicated his life to the service of science. He was influential in the real world dynamics, mathematician, astrophysicist, cosmologist, but specially he was a Science's Knight. He published works with legendary figures as F. Pirani \cite{bp}, D. Sciama \cite{bgs} and W. Rindler \cite{br}.}

-----------------------
\vskip 0.25cm
Es preciso y conveniente concluir esta primera parte haciendo referencia 
expl\'\i cita a una rese\~na que escribi\'o Bondi en {\it Nature} sobre el libro de S. Chandrasekhar titulado: {\it Verdad y Belleza: Est\'etica y Motivaciones en la Ciencia} \cite{chandra}, excelente libro por dem\'as:
\vskip 0.25cm
{\it LO BUENO, LO MALO Y LO FEO}

{\it Hermann Bondi}

{\it \textexclamdown Qu\'e libro tan espl\'endido! Su lectura es un placer, y para m\'{\i}, al menos,
la lectura continua se hizo compulsiva.

El hecho de que consista en un compendio de charlas (una de 1946, y las otras de distintas fechas entre 1975--1986)
significa que  cada una de las secciones son autocontenidas, y que adem\'as casi no hay solapamiento o repetici\'on.
Chandrasekhar es un distinguido astrof\'\i sico y cada una de sus charlas posee el sello de toda su obra:
precisi\'on, profundidad, lucidez. Lo que quiz\'as no vislumbr\'e fue la profunda formaci\'on hist\'orica 
del autor. Me apresuro en agregar que me result\'o claro que si Chadrasekhar decide hablar sobre asuntos
hist\'oricos o art\'\i sticos, su estudio ser\'a realizado a la perfecci\'on; lo que no result\'o obvio fueron los
tantos aspectos de estos temas que no figuran en sus conferencias... Conf\'\i o en que ning\'un lector de esta
rese\~na pueda pasar desapercibido mi entusiasmo por el libro, aun cuando debo tambi\'en hacer algunas acotaciones
cr\'\i ticas puesto que algunos aspectos del contenido no se ajustan al panorama que tengo de la ciencia. Primero, para
un leal disc\'\i pulo como yo de Popper, lo que hay sobre filosof\'\i a de la ciencia en el libro es incompleto, porque
no posee alg\'un tipo de perspectiva Popperiana. Segundo, el t\'\i tulo del libro me produjo picaz\'on (aunque,
afortunadamente para mi disfrute, finalmente esto result\'o irrelevante). Soy quiz\'as m\'as cauteloso que Popper en 
asignar verdad a la ciencia, y tambi\'en tengo poca f\'e en la utilidad de la belleza como gu\'\i a luminosa.
No s\'olamente me perturba la subjetividad en el reconocimiento de la belleza, sino que me parece  
confunde a la gente o emerge s\'olamente despu\'es de una contribuci\'on original. 
As\'{\i}, Chandrasekhar escribe bien y correctamente sobre la irrelevancia para la nueva ciencia de
buena parte de los \'ultimos trabajos de Eddington, de Einstein y de Milne. Estos trabajos fueron guiados en buena medida
y al extremo por consideraciones de belleza, especialmente belleza matem\'atica, y muy poco por la
dura y sucia f\'\i sica.

De nuevo, cuando Dirac, en su primera contribuci\'on excepcionalmente buena, produjo la ecuaci\'on
relativista del electr\'on, su desgarbo y fealdad fueron proclamadas. Posteriores trabajos, de
Dirac y otros, condujeron a nuevas formulaciones igualmente excepcionales en su belleza. De igual forma,
cuando surgieron los agujeros negros en la teor\'\i a, no hab\'\i a algo particularmente elegante
en las ecuaciones. Mucho despu\'es, el mismo Chadrasekhar, en su espl\'endido libro TEOR\'IA MATEM\'ATICA
DE LOS AGUJEROS NEGROS, compil\'o su propio trabajo y el de muchos otros para revelar los
patrones de una extraordinaria belleza. 

No digo que mostrar los descubrimientos originales de una forma concisa y mediante descripciones elegantes sea
considerablemente \'util para los que trabajar\'an posteriormente en el tema, sino que evidentemente esto no contribuy\'o
a los descubrimientos mismos. La belleza puede ser una excelente gu\'\i a en matem\'aticas, pero dudo que tenga
alg\'un valor en la f\'\i sica.

Esta exhibici\'on de ligera petulancia en mis puntos de vista, pos\'\i blemente heterodoxos, no deber\'\i a ser \'obice
para que cualquier lector potencial disfrute este excelente libro, como yo lo hice a cabalidad.}
\vskip 0.25cm
\textcolor{blue}{It is necessary and convenient to conclude this first part making an explicit reference to a note written by Bondi for {\it Nature} on the book of S. Chandrasekhar entitled: {\it Truth and beauty: Aesthetic motivations in Science} \cite{chandra}, an exceedingly good book:}
\vskip 0.25cm
\textcolor{blue}{\it THE GOOD, THE BAD AND THE UGLY}

\textcolor{blue}{\it Hermann Bondi}

\textcolor{blue}{\it What a splendid book! Reading it is a joy, and for me, at least, continuing reading became compulsive.}

\textcolor{blue}{\it The fact that this is a collection of lectures (one from 1946, the others bearing dates in the period 1975--1986) means that each of the sections is self--contained, yet there is almost no overlap or repetition. Chandrasekhar is a distinguished astrophysicist and every one of the lectures bears the hallmark of all his work: precision, thoroughness, lucidity.
What perhaps I had not foreseen is the depth of the author's historical scholarship. I hasten to add that is was willing to talk historical or artistic matters, his study would be thorough to perfection; what was not obvious was the that so many aspects of these subjects would figure in his lectures... I trust no reader of this review can have missed my enthusiasm for the book, yet I must also add some critical remarks because some aspects of the contents do not fit well into my picture of science. First, to a staunch disciple of Popper like myself, what philosophy of science there is in the book is incomplete, because it does not have any kind of Popperian perspective. Secondly, the very title of the book grates on me (though, fortunately for my enjoyment of it, this turns out to be irrelevant). I am perhaps even more cautious than Popper is assigning truth to science, and also have little faith in the usefulness of beauty, but it seems to me either to have misled people or to have arisen only well after the original contribution. Thus Chandrasekhar writes so well and so rightly about the irrelevance to later science of much of the later work of Eddington, of Einstein and of Milne. This work was guided far too much by considerations of beauty, and far too little by hard, dirty physics.}

\textcolor{blue}{\it Again, when Dirac, in his first outstanding contribution, produced the relativistic equation for the electron, its clumsiness and hence ugliness was pronounced. Later work, by Dirac and others, led to new mathematical formulations outstanding in their beauty. Equally, when black holes first arose in the theory, there was nothing particularly elegant about the equations. Much later, Chandrasekhar himself, in his splendid book THE MATHEMATICAL THEORY OF BLACK HOLES, put together his own work and that of many others to reveal a pattern of outstanding beauty.}

\textcolor{blue}{\it  I do not wish to deny that giving earlier discoveries such concise, elegant descriptions is enormously useful for subsequent workers, but evidently it did not contribute to the original discovery. Beauty may be an excellent guide in mathematics, but I doubt its value in physics.}

\textcolor{blue}{\it This slightly petulant display of my, perhaps heterodox, views should not stop any potential reader from enjoying this excellent book, as I did so fully.}

-----------------------
\vskip 0.25cm
He intentado, ojal\'a lo haya logrado, mostrar los signos de una personalidad singular. Ahora centrar\'e la atenci\'on en algunos 
 aspectos t\'ecnicos del genio y obra de Bondi, a trav\'es de dos art\'\i culos que representan una rica fuente de ideas y que se mantienen
 casi inc\'olumes despu\'es de cuarenta a\~nos. Tambi\'en intentar\'e revisar, sin profundidad, el trabajo que se ha realizado en Venezuela relacionado dir\'ectamamente con el de Bondi.
\vskip 0.25cm
\textcolor{blue}{I have attempted point out, hopefully, the signs of a peculiar personality. I will focus attention now on technical aspects of his genius and work through two papers, which represent a rich source of ideas and have remained standing after forty years. Also I will try a shallow review of the work done in Venezuela straightly connected with Bondi's work.}
 
-----------------------
\vskip 0.25cm

La obra de Bondi se puede clasificar en cuatro grandes \'areas: modelos cosmol\'ogicos;  modelos de acreci\'on; modelos esf\'ericos; radiaci\'on gravitacional.
El trabajo {\it Gravitational Waves in General Relativity. VII. Waves from Axi-Symmetric Isolated Systems} de H. Bondi, M. G. J. van de Burg y A. W. K. Metzner, publicado en {\it 
Proceedings of the Royal Society of London. Series A, Mathematical and Physical Sciences, Vo\-lume 269, Issue 1336, pp. 21--52, 1962}, es fundamental para la Rela\-tividad General. Aunque restringido a la simetr\'\i a axial y de reflexi\'on estableci\'o
las bases para la comprensi\'on de la radiaci\'on gravitacional fuera del contexto lineal. Inmediatamente se valor\'o este trabajo y fue gene\-ralizado por R. Sachs \cite{s62a} 
el mismo a\~no. Fue descubierta la exis\-tencia de un grupo de simetr\'\i a denominado ahora el Grupo de Bondi--Metzsner--Sachs (BMS) \cite{s62b} justamente a partir del trabajo de Bondi y en el intento de definir ener\-g\'\i a al menos en espaciotiempos asint\'oticamente planos. Bondi y colaboradores introdujeron una restricci\'on innecesaria al hacer una analog\'\i a con el electromagnetismo \cite{frau}. Pero, no es posible evitar radiaci\'on entrante mediante la condici\'on de Sommerfeld, debido al car\'acter no lineal de la Relatividad General. El tratamiento {\it conformal} del infinito resolvi\'o t\'ecnicamente estos ``detalles'' \cite{wald}. Se puede decir que el trabajo de Bondi--Sachs fue genera\-lizado por Winicour y Tamburino \cite{wt}. Gracias a ello es posible hoy en d\'\i a simular la radiaci\'on gravitacional proveniente de sistemas aislados, usando el calibre de Bondi. Debo se\~nalar que no ha sido posible construir un espaciotiempo glo\-balmente regular a partir del refe\-rencial de Bondi en el infinito--nulo--futuro. Pero tampoco hubiera sido posible el desarrollo de los c\'odigos en la formulaci\'on caracter\'\i stica sin la piedra preciosa de 1962. 
\vskip 0.25cm
\textcolor{blue}{The entire work of Bondi can be classified in four wide areas: cosmological models; accretion models; spherical models; gravitational radiation. The work {\it Gravitational Waves in General Relativity. VII. Waves from Axi-Symmetric Isolated Systems} by H. Bondi, M. G. J. van de Burg y A. W. K. Metzner, published in {\it Proceedings of the Royal Society of London. Series A, Mathematical and Physical Sciences, Vo\-lume 269, Issue 1336, pp. 21--52, 1962}, is fundamental for General Relativity. Although restricted to axial and reflexion symmetries this paper is a cornerstone for a comprehensive view of gravitational radiation off linear regime. The impact was immediate and generalized by R. Sachs \cite{s62a} the same year. It was discovered the existence of the symmetry group called now the Bondi--Metzsner--Sachs (BMS) \cite{s62b} just after Bondi's work trying to define energy at least in asymptotically flat spacetimes. Bondi and collaborators introduced an unnecessary restriction following an electromagnetic analogy \cite{frau}. But, it is not possible to avoid incoming radiation by means of Sommerfeld's condition, because of the non linear character of General Relativity. The conformal treatment of infinity technically solved these ``details'' \cite{wald}. It can be said that the Bondi--Sachs work was extended by Winicour and Tamburino \cite{wt}. For this reason it is possible nowadays the simulation of gravitational radiation from isolated sources, using the gauge of Bondi. Up to now it has been not possible to construct a globally regular spacetime from a Bondi's referential at future--null--infinity. But nor would have been possible to develop numerical codes in the characteristic formulation without the 1962's gem.}

-----------------------
\vskip 0.25cm
Ahora bien, resulta curioso que el trabajo de Bondi {\it The contraction
 of gravitating spheres} publicado en {\it Proceedings of the Royal Society of London, Series A, Vo\-lume 281, Number 1384 / August 25, 1964} haya recibido atenci\'on pr\'acticamente s\'olo desde Venezuela. Veamos. En este trabajo Bondi estudia el problema de una distribuci\'on material esf\'erica introduciendo la noci\'on de observadores com\'oviles con el fluido y localmente minkowskianos. Esto permite definir las variables f\'\i sicas y en consecuencia una des\-cripci\'on transparente de la din\'amica. Aunque el tratamiento es en un referencial no com\'ovil, la definici\'on de variables f\'\i sicas para observadores apropiados excluye toda posible
ambig\"uedad en la interpretaci\'on de los resultados. La evoluci\'on lenta es posible desde su enfoque y extrae ciertos resultados generales para un fluido adiab\'atico, haciendo conside\-raciones f\'\i sicas razonables. Bondi critica la idea innecesaria de suponer adicionalmente una ley de conservaci\'on de bariones, ya que en Relatividad General la noci\'on de part\'\i culas no existe. Hasta donde tenemos conocimiento esta afirmaci\'on de Bondi no fue rebatida. Extra\~na adem\'as que el tratamiento de observadores com\'oviles en el local minkowskiano no sea el tratamiento est\'andar para incorporar materia en Relatividad General. Por si fuera poco, en una segunda parte del art\'\i culo se adaptan las coordenadas de radiaci\'on para estudiar problemas no adiab\'aticos o de radiaci\'on profusa, acerc\'andose m\'as a una situaci\'on realista y plantea un posible esquema para la obtenci\'on de modelos.
\vskip 0.25cm
\textcolor{blue}{Now, it is curious how the work of Bondi {\it The contraction
 of gravitating spheres} published in {\it Proceedings of the Royal Society of London, Series A, Vo\-lume 281, Number 1384 / August 25, 1964} received almost exclusive attention only from Venezuela. Let's see. Bondi studied in this work the problem of a spherical matter distribution introducing the notion of comoving observers with fluid and locally Minkowskians. It allows to define the physical variables and consequently a clear description of dynamics. Although the treatment is in noncomoving coordinates, the definition of physical variables for proper observers makes unambiguous the interpretation of results. Slow motion is possible from his approach, extracting some general results under reasonable physical assumptions for an adiabatic evolution. Bondi criticizes the idea of an additional law of baryons conservation due to the unnecessary  notion of particles in General Relativity as a continuum theory. As far we know this assertion was not debated. It is surprising that the approach of comoving and local Minkowskian observers is not the standard to include matter in General Relativity. If not enough, in the second part of the paper, the radiation coordinates are adapted to study non adiabatic or profusely radiative systems. He describes more realistic situations and proposes a possible sketch to get models.}
 
-----------------------
\vskip 0.25cm

Dieciseis a\~nos m\'as tarde Herrera, Jim\'enez y Ruggeri \cite{hjr} proponen un m\'etodo seminum\'erico que retomaba las ideas originarias en este trabajo de Bondi y que posteriormente se interpretar\'\i a como la aproximaci\'on post-casi-est\'atica \cite{hbds02}. En una revisi\'on bibliogr\'afica profunda pudimos constatar que con excepci\'on de los relativistas venezolanos, el trabajo de 1964 ha permanecido ``dormido'' inexplicablemente, en lo referente al tratamiento de la materia. Su extensi\'on, uso y difusi\'on se debe a Luis Herrera y colaboradores. El tratamiento de materia en la formulaci\'on caracter\'\i stica est\'andar de la rela\-tividad num\'erica no usa los observadores com\'oviles en el local minkowskiano de Bondi. Ni siquiera en la versi\'on actual m\'as general y poderosa, como he\-rramienta, de Jos\'e Font y Philippos Papadoupolos \cite{fp}.
\vskip 0.25cm
\textcolor{blue}{Sixteen years later Herrera, Jim\'enez and Ruggeri \cite{hjr} patterned a seminumerical method originally envisioned by Bondi and recently interpreted as the post--quasi--static approximation \cite{hbds02}. Doing an exhaustive bibliographic revision we check that with exception of Venezuelan physicists the 1964's work has remained ``dormant'' without explanation, concerning the matter treatment. Its extension, use and diffusion is owing to Luis Herrera and collaborators. The treatment of matter in the standard characteristic formulation of numerical relativity does not use, at least explicitly, local Minkowskian comoving observers of Bondi. It can be corroborated in the most general and powerful current version, as a tool, of Jos\'e Font and Philippos Papadoupolos \cite{fp}.}

-----------------------
\vskip 0.25cm

Es importante se\~nalar que Demetrios Christodoulou \cite{chris}, con prop\'ositos anal\'\i ticos,
Dalia Goldwirth y Tsvi Piran \cite{gp}, Roberto G\'omez y Jeffrey Winicour \cite{gw}, con prop\'ositos num\'ericos, usaron la m\'etrica de Bondi (de radiaci\'on)
bajo simetr\'\i a esf\'erica en el contexto del colapso de un campo escalar autogravitante.  Este sistema es el m\'as sencillo posible para estudiar el colapso gravitacional, as\'{\i} como el sistema del trabajo de 1962 es el m\'as sencillo para estudiar radiaci\'on gravitacional.
 Aunque idealizado y con poco, si acaso ning\'un, contenido
f\'\i sico, condujo al descubrimiento del comportamiento cr\'\i tico \cite{choptuik}. Es interesante observar que  a la
vez el problema del campo escalar sin masa, esto es, la radiaci\'on escalar,  guarda estrecha relaci\'on
con el problema de la radiaci\'on gravitacional, magistralmente estudiado por Bondi y colaboradores en el trabajo de 1962.      
\vskip 0.25cm
\textcolor{blue}{It is important to point out that Demetrios Christodoulou \cite{chris}, with analytical purposes, Dalia Goldwirth and Tsvi Piran \cite{gp}, Roberto G\'omez and Jeffrey Winicour \cite{gw}, with practical purposes, use Bondi's metric (of radiation) under spherical symmetry in the context of a self--gravitating scalar field. This system is the simplest to study the gravitational collapse,
as the system of 1962's work is the simplest to study gravitational radiation. Although idealized and with little, if none, physical content, led to the discovery of critical behavior \cite{choptuik}. It is interesting to note that at the same time the problem of the masslees
scalar field is closely related with the gravitational radiation problem, masterly studied by Bondi and collaborators in his 1962's work.}

-----------------------
\vskip 0.25cm
La orden del Caballero Bondi es continuar con su ho\-norable tradici\'on. Es explorar por qu\'e su trabajo de 1964 no ha sido
valorado entre los relativistas num\'ericos, en lo referente al tratamiento de la materia.
 Nos preguntamos: ?`por qu\'e el enfoque de los observadores com\'oviles localmente minkowskianos de Bondi no es el est\'andar en Relatividad Num\'erica?
 ?`ser\'a que la heterodoxia no favorece a la industria de estrellas en el firmamento?
 
\vskip 0.25cm 

\textcolor{blue}{The Knight Bondi's order is to continue with his honorable tradition. It is to explore why his 1964's work has not been valued among numerical relativity practitioners in the treatment of matter issues.  Why does the approach of Bondi to treat with matter is not the standard in numerical relativity? should heterodox views go against production of heaven stars?}

\section*{Agradecimientos} El autor agradece a Luis Herrera, Alicia Di Prisco, Jeffrey Winicour, Luis Rosales y Carlos Peralta por motivar la presentaci\'on de esta semblanza de Bondi y por sus comentarios.
\textcolor{blue}{\section*{Aknowlegedments} Thanks to Luis Herrera, Alicia Di Prisco, Jeffrey Winicour, Luis Rosales and Carlos Peralta for motivating me to write this biographical sketch of Bondi; also for their comments.}

\end{document}